\documentclass{article}
\usepackage{spconf,amsmath,graphicx}

\usepackage{enumitem}
\setlist{nosep, leftmargin=14pt}

\usepackage{mwe} 

\title{No-reference denoising of low-dose CT projections}
\name{Elvira Zainulina$^{\ddag\dagger}$, Alexey Chernyavskiy$^{\dagger}$, Dmitry V. Dylov$^{\ddag}$\thanks{Correspondence to: Elvira Zainulina -- elvira.zainulina@philips.com}}

\address{ $^{\dagger}$Philips AI Research, $^{\ddag}$ Skolkovo Institute of Science and
Technology, Moscow, Russia}
\begin{document}
%
\maketitle
\begin{abstract}
Low-dose computed tomography (LDCT) became a clear trend in radiology with an aspiration to refrain from delivering excessive X-ray radiation to the patients. 
The reduction of the radiation dose decreases the risks to the patients but raises the noise level, affecting the quality of the images and their ultimate diagnostic value.
One mitigation option is to consider pairs of low-dose and high-dose CT projections to train a denoising model using deep learning algorithms; however, such pairs are rarely available in practice. 
In this paper, we present a new self-supervised method for CT denoising. 
Unlike existing self-supervised approaches, the proposed method requires only noisy CT projections and exploits the connections between adjacent images. 
The experiments carried out on an LDCT dataset demonstrate that our method is almost as accurate as the supervised approach, while also outperforming several modern self-supervised denoising methods. 
\end{abstract}
\section{Introduction}
\label{sec:intro}

The potential health risks for patients caused by X-ray radiation of computed tomography (CT) have led to the emergence of the low-dose CT~\cite{LDCT_cancer}. Although the reduction of radiation dose decreases the risks, it also increases the level of noise. The deterioration of the quality in CT projections leads to streaks and blurriness in the reconstructed CT slices. While there have been progress in applying convolutional neural networks (CNNs) for denoising the reconstructed CT slices~\cite{CPCE-3D},
the raw projections contain much more information which could and should be used for increasing the signal-to-noise ratio and to contribute to more accurate reconstruction.  

Traditional denoising algorithms rely on using regularization with special penalty functions such as total variation. These algorithms often require careful case-by-case tuning of parameters. On the other hand, the most recent algorithms based on deep learning (DL) are supervised, i.e. they require pairs of low-noise and high-noise images of the same scene for training. In a clinical setting, for example in a radiology suite, the acquisition of such paired data can take at least twice longer than the regular CT exam and would increase exposure to the X-ray dose. Generating the noisy data artificially is an option to alleviate the lack of the image pairs, but the addition of the noise to the images, especially the medical ones, can bias the predictions of a CNN because no synthetic noise can completely emulate the real one to portray all physical phenomena encountered in a CT machine. 

Recently, self-supervised methods for reference-free image denoising have been proposed. In these methods, the noiseless signal is deduced from the noisy image itself. Research has shown that self-supervised methods produce images with quality comparable to that of images denoised by the supervised methods without requiring 
reference images. 

In~\cite{Noise2Noise}, Lehtinen et al. have proposed a CNN that considers pairs of noisy images of the same scene with two independent realizations of noise. Given the first image, the CNN is trained to produce the second image, and learns an approximation to the noise-free image as a result. This Noise2Noise method was applied for denoising X-ray projections and CT images and compared to a supervised model in~\cite{n2n_ct2020}. It gave acceptable results but the images were over-smoothed.
Unless there is a way of getting two co-registered images of the same scene, e.g. from a still camera, this approach is not practical. Most often, only one noisy image of a particular scene is available. Custom addition of noise to images before applying this approach is also questionable since the mismatch between the statistical properties of the synthetic data and those of the real data may lead to a decrease in image quality at test time compared to the performance during training.

Other current self-supervised denoising methods require only noisy images, but they add complexity to the model.
In~\cite{Noise2Void}, Krull et al. have described Noise2Void, an algorithm that uses pixel masking
during the training process making it less efficient. 
A solution to overcome this drawback was proposed by Laine et al. in~\cite{Noise2Void_NVIDIA}. They developed an architecture based on the U-Net~\cite{ronneberger2015u} that includes special convolution and downsampling layers. This architecture allows to obtain a blind spot at the center of the patch, instead of masking. Also, the authors of~\cite{Noise2Void_NVIDIA} introduced loss functions
that take into account Gaussian or Poisson distribution of the noise. However, their method requires passing one image through a CNN four times for removing noise or making a model four times bigger. In~\cite{st_n2v}, a more complicated statistical approach was applied for denoising CT images but it was applied only to reconstructed scans and sinograms, not CT projections. 

To our knowledge, none of the existent self-supervised reference-free denoising methods use information contained in sequences of images. Each image is denoised separately, which leads to sub-optimal image quality. 
We propose Noise2NoiseTD, a time-distributed method that exploits the information redundancy in sequences of images such as CT projections. It requires neither pairs of low-noise and high-noise images nor pairs of high-noise images of the same scene. It makes the denoising process simpler compared to existent DL-based self-supervised approaches. 

\section{Methodology}
\label{sec:methodology}
\subsection{Noise2NoiseTD approach overview}
\label{sec:overview}
The proposed approach is based on the assumption that, given a sequence of $2k+1$ image frames (CT-projections) $p(\theta\pm i\Delta\theta),\; i=1,\dots,k$, where $\theta$ is the angle of rotation of the X-ray source around the patient, the content of the frames can be distinguished from the noise using similarities found among them by a neural network. Since the noise is independent on each CT projection, the features extracted from the sequence of frames will reflect only the information about the projected structures (the anatomy) and allow to recover the noise-free projection $p(\theta)$. The choice of the number of the adjacent frames $k$ depends on how much the content of the frames overlaps, and the computational and memory capacity of the device. 

We propose to use the bidirectional convolutional memory units (Bi-ConvLSTM) for carrying information about the adjacent images.  These units allow to extract the features corresponding to the slight consequent change of the structures that are observed from the first to the last viewing angle, and vice versa, and then combine these features. Thus, Bi-ConvLSTM units provide a stable restoration of the middle frame in the sequence that is being denoised.  
These units work with features that are extracted from each frame independently by some CNN. Finally, the features extracted from the frames of the sequence, with the exception of the middle projection, are summed up and processed by another CNN. The task of this network is to process and fuse the results to obtain the denoised middle projection; therefore it can have a simpler architecture. 

Since no noiseless ground truth is available, the training of Noise2NoiseTD proceeds in a self-supervised way. We use a no-reference loss function that is based on the loss function proposed in~\cite{Noise2Void_NVIDIA}. 
The network architecture and the loss function will be described in detail in the next subsections.

\subsection{Network architecture}
\label{sec:architecture}
\begin{figure}
\begin{center}
    \includegraphics[width=1.0\linewidth]{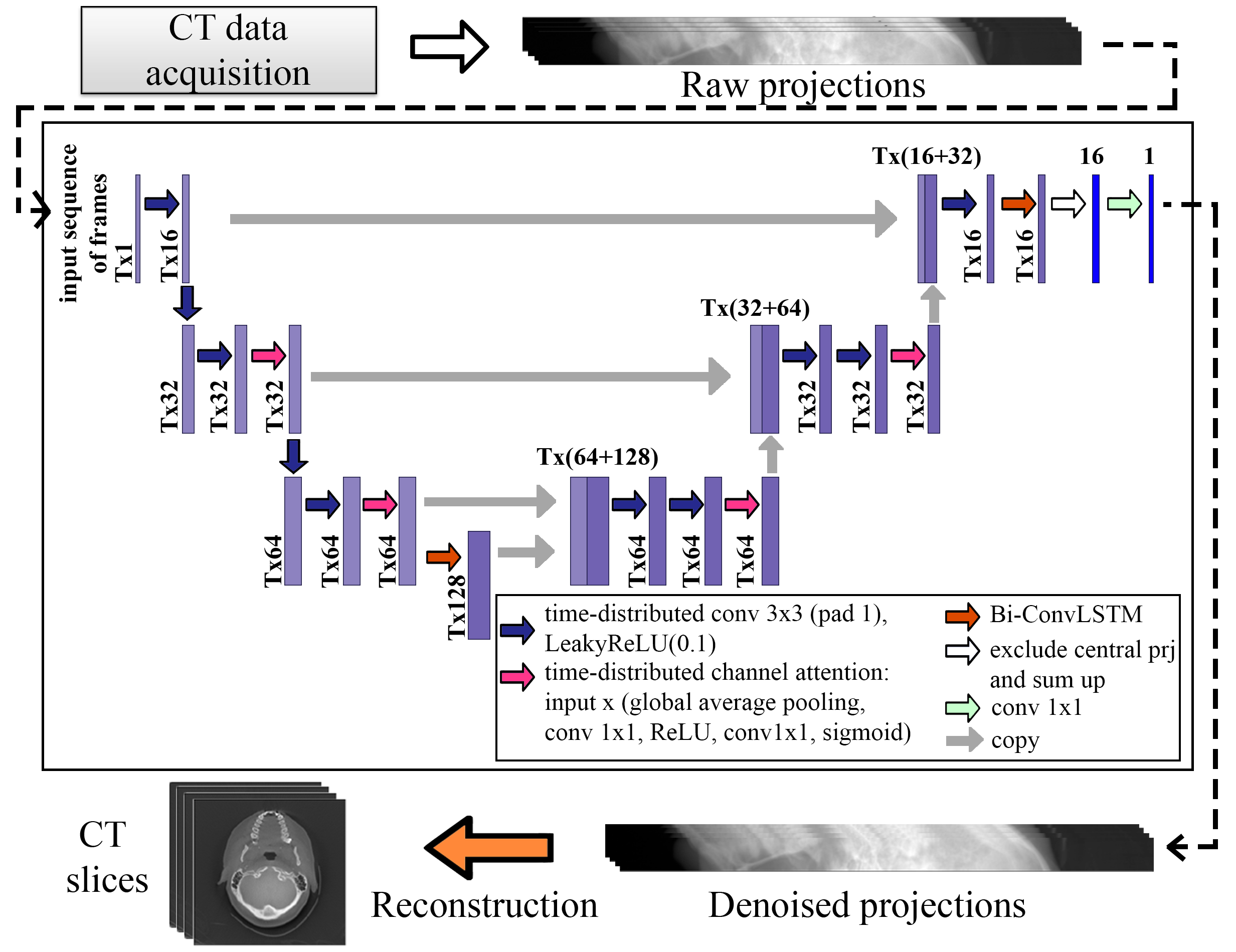}
\end{center}
  \caption{Architecture of the Noise2NoiseTD model.}
\label{n2n_model}
\end{figure}
Figure~\ref{n2n_model} shows the proposed Noise2NoiseTD model. The backbone of our implementation of the Noise2Noise approach is based on the U-Net architecture, except that it does not have any pooling layers and all levels preserve the width and height of the input tensors using replicate padding. Since some features can contain more information related to the anatomical structures, and others may contain more information about the corruptions, we used a feature attention mechanism. Channel attention allows to focus on more informative features. 
Channel attention layers~\cite{SE} are added after each block that is on the second or third level of the model. 

To account for connections between subsequent frames, Bi-ConvLSTM blocks are included to the model. These blocks are based on the ConvLSTM layers introduced in~\cite{weather}. We do not use past cell status for gates calculation; it reduces the number of model parameters without really affecting image quality. A Bi-ConvLSTM layer consists of two ConvLSTM-cells that process image frames in two opposite directions. The forward and backward outputs are combined by a summation operator. As suggested in~\cite{sequential}, the first Bi-ConvLSTM layer is inserted into the bottleneck and the second one is plugged into the end of the network.

The output of the last Bi-ConvLSTM layer is summed up along the time axis, as we want to predict only the middle projection in the sequence. With the purpose of preventing overfitting to the noisy middle projection, we make the network ``blind'' to it by excluding the corresponding Bi-ConvLSTM output from the summation. Next, the aggregated output is processed by a $1\times1$ convolution in order to obtain the proper number of output channels. Following~\cite{bias-free}, all the convolutions in the model are made bias-free.

\subsection{Loss function}
\label{sec:loss}
It was stated in~\cite{DIWAKAR201873} and~\cite{noise_prop} that the noise in CT projections is more likely to be Gaussian. However, for low-dose CT the noise at each pixel can be more accurately modeled as an independent random variable sampled from a mixed Poisson-Gaussian distribution. We found from our experiments that this assumption gave  better results than when the noise was assumed to be purely Gaussian or Poisson. Based on this assumption, one can derive a reference-free loss function suitable for self-supervised model training that accounts for the noise distribution.

According to~\cite{Noise2Void_NVIDIA}, the distribution of the noisy data $\mathbf{y}=\left(y_1,\dots,y_n\right)$ given its neighbourhood $\mathbf{\Omega_y}=\left(\Omega_{y_1},\dots,\Omega_{y_n}\right)$ relates to the distribution of the clean data $\mathbf{x}=\left(x_1,\dots,x_n\right)$ in the following way:
\[\underbrace{p(y_i|\Omega_{y_i})}_{\text{Noisy\; observation}} = \int \underbrace{p(y_i|x_i)}_{\text{Noise\; model}}\underbrace{p(x_i|\Omega_{y_i})}_{\text{Clean\; prior}}dx_i.\]
As only noisy data is available, the network that models prior distribution can be trained by minimizing the following log-likelihood function:
\[\mathcal{L}=-\sum_{i=1}^{n}\log{p(y_i|\Omega_{y_i})}.\]
For predicting $\mathbf{x}$, in addition to $\mathbf{\Omega_y}$, $\mathbf{y}$ can be included using Bayes rule:
\[p(x_i|y_i,\Omega_{y_i})\propto p(y_i|x_i)p(x_i|\Omega_{y_i}).\]

The algorithm for training a CNN with noisy data is the following~\cite{Noise2Void_NVIDIA} (for convenience we omit the index i):
\begin{enumerate}
    \item Train the model to map the pixels in the patch into the mean $\mu_x$ and standard deviation $\sigma_x$ of the Gaussian approximation of the distribution of the clean data $p(x|\Omega_y)$.
    \item During the test phase obtain $\mu_x$ and $\sigma_x$ using the trained network. Then compute $\mathrm{E}_x[p(x|y,\Omega_y)]$.
\end{enumerate}

Assuming a mixed Poisson-Gaussian distribution, 
\[y=Poisson(\lambda x)/\lambda + \mathcal{N}(0, a), \]
where $\lambda$ is the maximum event count and $a$ is the variance of the additive Gaussian noise.
After approximation, the noisy data is modeled as $y=\mathcal{N}(\mu_x, \sigma_x^2 + \mu_x/\lambda + a)$.
$a$ and $\lambda$ are the unknown parameters that are learned together with the main model. The variance of the noise is $\sigma_n^2=\mu_x/\lambda + a$.

The loss function for CNN training is therefore given as:
\begin{equation}
   \mathcal{L}=\sum_i\left( \frac{(y_i-\mu_{i})^2}{2\sigma_i^2}+\frac{1}{2}\log{\sigma_i^2}-0.1\sigma_i\right), 
\label{eqn:loss}
\end{equation}
where $-0.1\sigma_i$ is a regularization parameter proposed in~\cite{Noise2Void_NVIDIA} for the case of unknown noise parameters that encourages explaining the observed noise as corruption instead of uncertainty about the clean signal. The posterior mean estimate, i.e. the prediction of the denoised image, is computed as follows:
\begin{equation}
    \mathrm{E}_x[p(x|y,\Omega_y)] = \frac{\mu_x\sigma_n^2+y\sigma^2_x}{\sigma_n^2+\sigma^2_x}.
\end{equation}

\subsection{Dataset}
\label{sec:dataset}
The experiments were carried out on the projection data from the LDCT dataset~\cite{LDCT} published by the Mayo clinic. CT projection data is provided for both full and simulated lower dose ($25\%$ of the routine dose) levels. Since the dose levels are different for the head and the abdomen, we trained two independent models for denoising the corresponding projections. They were also tested independently. For our experiments we selected sets of CT projections coming from seven randomly chosen patients: five containing head data, and two consisting of abdomen projections. 

For abdominal projections, we used data from one patient and randomly selected $21800$ and $3200$ projections for the train and validation sets respectively, in such a way that at least $100$ projections were adjacent. We used the first $6000$ substantive adjacent projections of the other patient with abdomen projections as test set; we denote it A-test.

For models that denoise head projections, the train, validation and test datasets consisted of $75\%$, $12.5\%$, $12.5\%$ correspondingly, of the data from 
three patients ($23616$ projections in total). We call the test part of this data  H-test. Since the acquisition geometry for these projections is axial, we split the data so that the projections from the same acquisition circle were included into one dataset. We also tested the models on full data from the two other patients (patient-$1$ and patient-$2$) in order to check the generalizability of the models and calculate metrics on the full set of projections.

\section{Experiments and results}
\label{sec:results}

\begin{figure}
\begin{minipage}[b]{0.48\linewidth}
  \centering
  \centerline{\includegraphics[width=4.5cm]{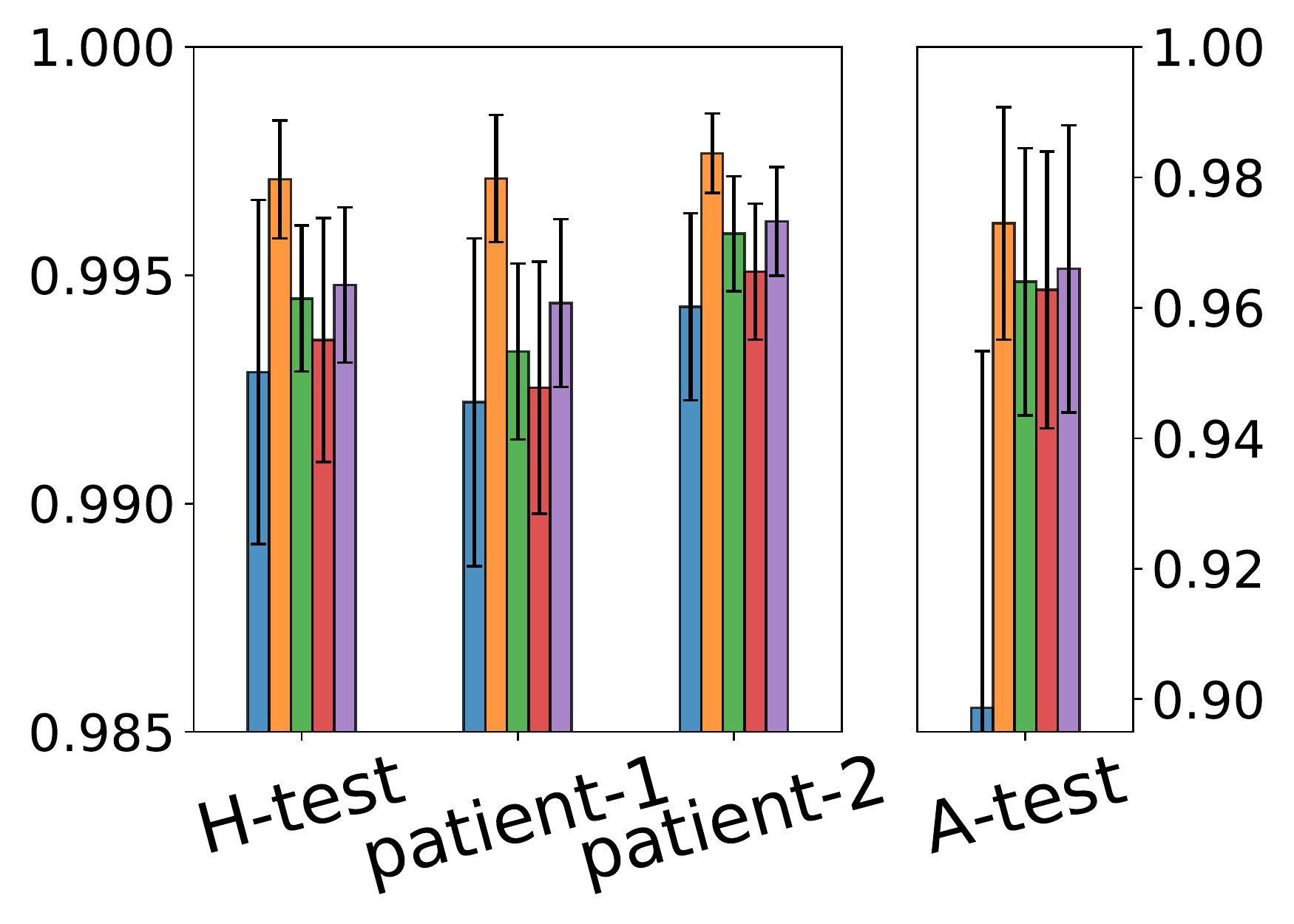}}
  \centerline{(a) projection domain}\medskip
\end{minipage}
\hfill
\begin{minipage}[b]{0.48\linewidth}
  \centering
  \centerline{\includegraphics[width=4.5cm]{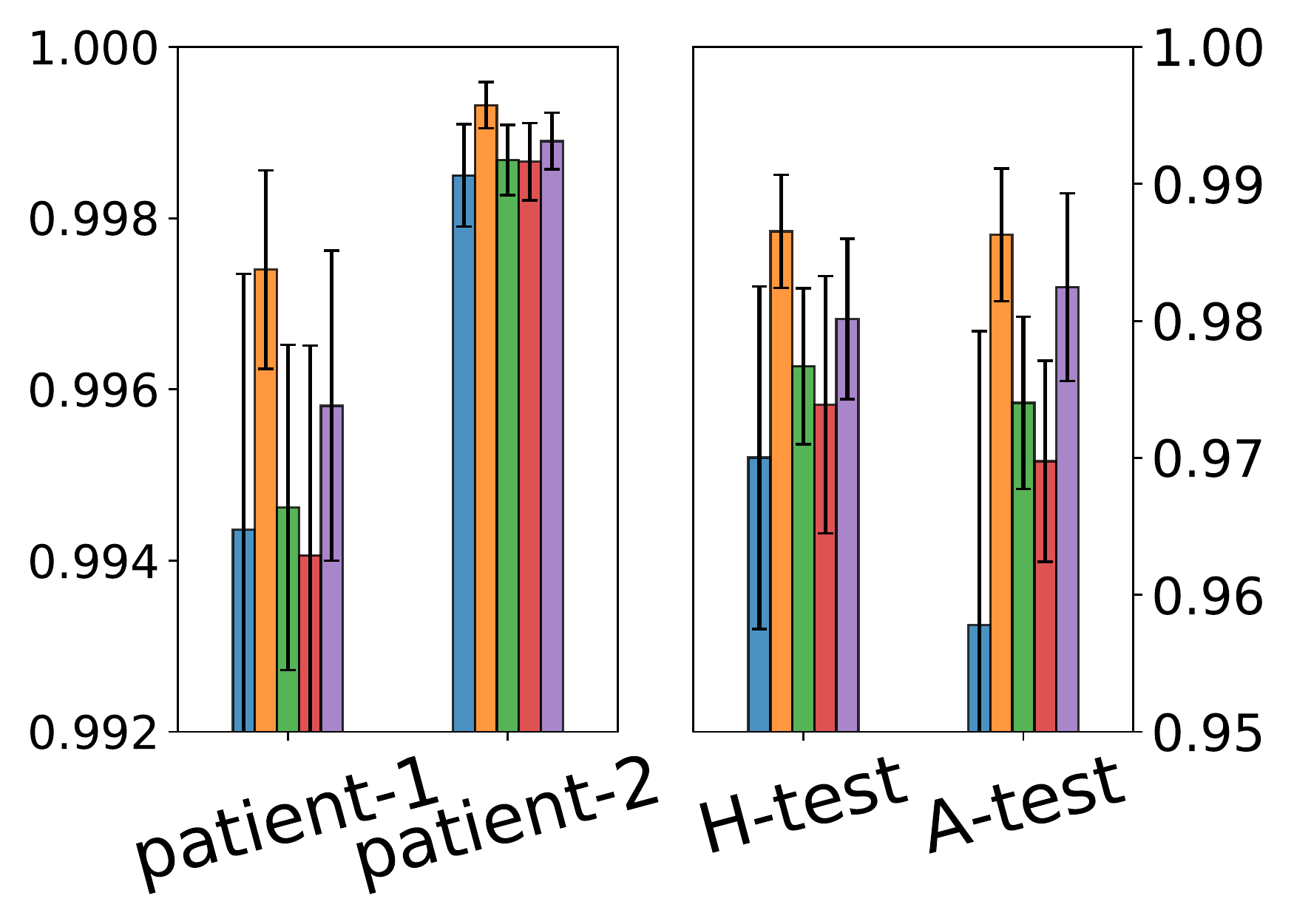}}
  \centerline{(b) image domain}\medskip
\end{minipage}
\begin{minipage}[b]{1.0\linewidth}
  \centering
  \centerline{\includegraphics[width=6.5cm]{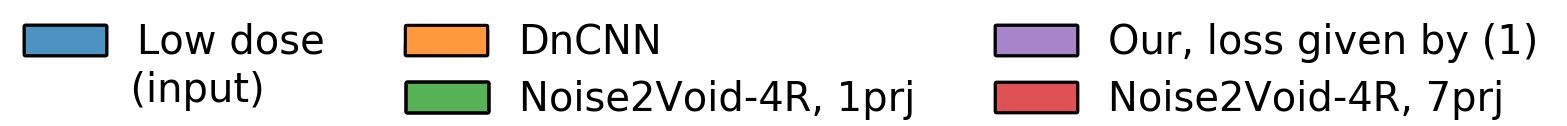}}
\end{minipage}

\caption{SSIM between the denoised and the full-dose images.}
\label{fig:ssim}
\end{figure}

Besides using the loss in (\ref{eqn:loss}), we also trained our model with the MSE loss. In this case we computed the loss between the middle projection and its denoised version. We chose to compare results against the recent self-supervised approach~of~\cite{Noise2Void_NVIDIA}. We observed that the algorithm of~\cite{Noise2Void_NVIDIA} produced results of higher quality when its original backbone CNN was replaced by DnCNN~\cite{zhang2017beyond}, so we decided to compare our approach with this modification of~\cite{Noise2Void_NVIDIA} that we call Noise2Void-$4$R ($4$R stands for the four rotations used to denoise an image patch). Finally, we performed denoising using DnCNN trained in supervised mode. 
Because our approach uses adjacent projections, in order to make the comparison more fair, we included adjacent projections as inputs to the Noise2Void-$4$R and DnCNN as additional channels.
A reasonable trade-off between complexity and level of detail for these models was achieved by taking three adjacent projections from each side, so the input consisted of seven projections.

\begin{figure}
\begin{minipage}[b]{1.0\linewidth}
  \centering
  \centerline{\includegraphics[width=8cm]{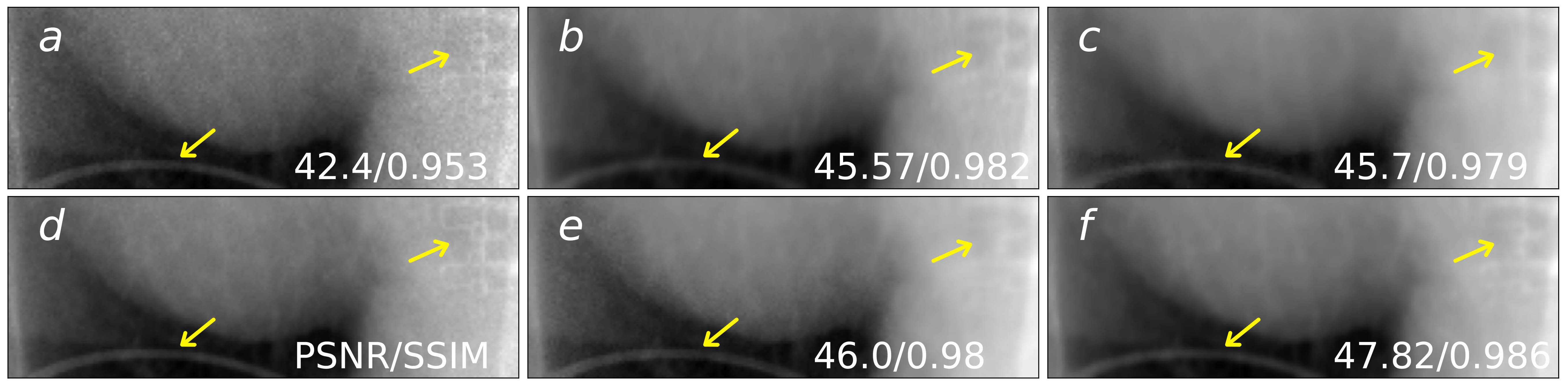}}
  \centerline{(1) Fragment of abdominal CT projection (A-test)}\medskip
\end{minipage}

\begin{minipage}[b]{1.0\linewidth}
  \centering
  \centerline{\includegraphics[width=8cm]{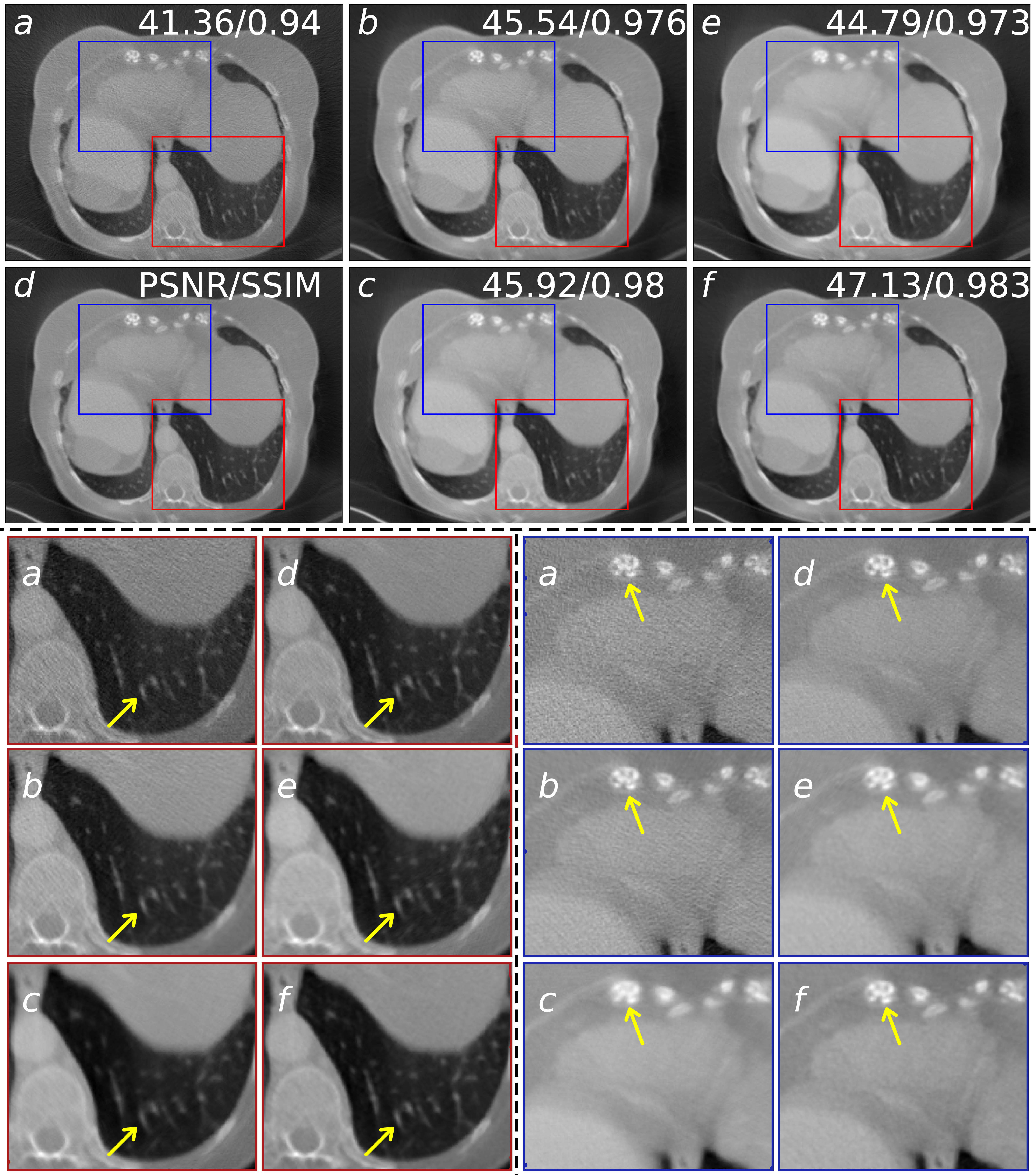}}
  \centerline{(2) Abdominal CT scan (A-test)}\medskip
\end{minipage}

\caption{Fragments of (1) CT projections and (2) CT scans reconstructed from original projections: (a) low-dose, (d) full-dose; from projections denoised by self-supervised models: (b) our model, MSE loss, (c) Noise2Void-$4$R, using $1$ projection, (e) our model, loss given by (\ref{eqn:loss}); from projections denoised by DnCNN (supervised): (f) using $7$ projections.}
\label{fig:ims}
\end{figure}

All self-supervised models were trained in Pytorch using Adam optimizer with default parameters, learning rate $10^{-4}$, and minibatch size of $8$. 
The noise model (parameters $a$ and $\lambda$) was trained together with the main denoising model.
The minibatches consisted of random $64\times64$ crops. We assumed that full-dose projections were available for the validation sets, and used them only to decide when to stop training. Namely, we trained our models until the moment when the PSNR began to decrease and $0.86\cdot\text{SSIM} + 0.14\cdot\text{L1}$-loss~\cite{losses} began to grow on the validation set.  

Supervised models were trained using MSE loss, Adam optimizer with default parameters, learning rate $4\cdot 10^{-4}$ and minibatch size of $64$. The minibatches were formed as in the case of the self-supervised training. The training was performed until the moment the training curve reached a plateau.

We compared the denoising results in the projection domain, and also in the image domain that is usually of greater diagnostic and practical interest. We used the TIGRE toolbox~\cite{TIGRE} to reconstruct the CT slices from the projections. We computed the SSIM metric between high-dose projections and, first, the low-dose projections and, next, the denoised projections (we did the same with the corresponding CT slices). 
The SSIM metrics are shown in Figure~\ref{fig:ssim}.
The effect of our denoising is more pronounced for abdomen data, probably because the dose (and the image quality) for head CT is higher than for abdomen area. Interestingly, for Noise2Void-4R the use of several frames resulted in a smaller improvement of SSIM than when using only one frame.

We also evaluated the models visually, as image similarity metrics are not always demonstrative. The reconstructed CT scans are shown in Figure~\ref{fig:ims}. The results demonstrate that our model performs better denoising than the Noise2Void-$4$R model, and the image quality is comparable with the supervised DnCNN model. Whereas the Noise2Void-$4$R model produced overly smooth reconstructions even if it used several adjacent projections as input, our model was better at preserving edges and fine details.

\section{Conclusion and discussions}
\label{sec:conclusion}
We demonstrated a new approach for denoising CT projections that supports training the model in the self-supervised mode and allows to denoise sequences of images depending only on the features extracted from these sequences. We compared our Noise2NoiseTD model to the state-of-the-art self-supervised denoising model~\cite{Noise2Void_NVIDIA} and to the popular supervised DnCNN network~\cite{zhang2017beyond}. Our model outperformed the self-supervised denoising model, and although we did not use high quality ground truth images during training, it produced results that are comparable to those of the supervised model.

Since the experiments were carried out on simulated data, the method should be tested on real data as well. Also, we plan to refine the learning process to consider projection normalization for more accurate noise approximation and estimation of its parameters. The image quality metrics that we used in this work may not correlate well with the opinion of radiologists. A qualitative comparison of denoising methods should be carried out with the help of domain experts. 

Because our method is not specific to the imaging modality 
except for the chosen noise model, it can be adopted for denoising sequences of other medical images, such as dynamic PET, dynamic MR, spectral CT, ultrasonograms, etc.

\section{Compliance with Ethical Standards}
This research study was conducted retrospectively using human subject data made available in open access by The Cancer Imaging Archive (TCIA). Ethical approval was not required as confirmed by the license attached with the open access data.

\section{Acknowledgments}
The authors declare no conflicts of interest.

\bibliographystyle{IEEEbib}
\bibliography{strings,refs}

\end{document}